\documentclass[journal] {IEEEtran}

\usepackage{epsfig}
\usepackage{amsmath}
\usepackage{subfigure}

\begin{document}

\def\e{\begin{equation}}
\def\f{\end{equation}}
\def\ea{\begin{eqnarray}}
\def\fa{\end{eqnarray}}
\def\=#1{\overline{\overline{#1}}}
\def\_#1{{\bf #1}}
\def\o{\omega}
\def\E{\epsilon}
\def\M{\mu}
\def\.{\cdot}
\def\x{\times}
\def\no{\noindent}
\def\l#1{\label{eq:#1}}
\def\r#1{(\ref{eq:#1})}
\def\d{\nabla}
\def\D{\partial}
\def\b{\delta}
\def\ds{\displaystyle}

\def\l#1{\label{eq:#1}}
\def\r#1{(\ref{eq:#1})}

\renewcommand{\Re}{\mathop{\rm Re}\nolimits}
\renewcommand{\Im}{\mathop{\rm Im}\nolimits}

\title{Subwavelength resolution with three-dimensional isotropic transmission-line lenses}

\author{Pekka~Alitalo and Sergei~A.~Tretyakov,~\IEEEmembership{Senior Member,~IEEE}
\thanks{P.~Alitalo and S.~A.~Tretyakov are with TKK, Helsinki University of Technology -- SMARAD-Radio Laboratory, P.O. Box 3000, FI-02015 TKK, Finland. (e-mail: pekka.alitalo@tkk.fi).}}

\maketitle

\begin{abstract}

Dispersion, impedance matching and resolution characteristics of
an isotropic three-dimensional flat lens (``superlens'') are
studied. The lens is based on cubic meshes of interconnected
transmission lines and bulk loads. We study a practical
realization of the lens, based on the microstrip technology. The
dispersion equations that have been previously derived, are
verified with full-wave simulations. The isotropy of the structure
is verified with analytical as well as simulation results. The
resolution characteristics of a practically realizable, lossy lens
are studied analytically.

\end{abstract}


\section{Introduction}

Materials with simultaneously negative material parameters
(double-negative or backward-wave materials, where permittivity
$\varepsilon$ and permeability $\mu$ are both effectively
negative)~\cite{Veselago} have received a lot of interest in the
recent literature. One of the most exciting applications of these
materials is a device capable of subwavelength resolution
(resolution that exceeds the diffraction limit)~\cite{Pendry}. The
first demonstrations of realized artificial backward-wave
materials were done in the microwave region using periodic
structures consisting of metal wires (negative $\varepsilon$) and
split-ring resonators (negative $\mu$)~\cite{Shelby}.

Also the use of loaded transmission-line networks has been
proposed for the realization of wide-band and low-loss
backward-wave materials in the microwave
region~\cite{Caloz,Eleftheriades}. These networks are inherently
one-or two-dimensional structures~\cite{Caloz2,Grbic}. Recently,
also three-dimensional, isotropic transmission-line-based
backward-wave materials have been
proposed~\cite{Grbic3,Hoefer,Alitalo1} and
realized~\cite{Alitalo2}.

It has been shown that subwavelength imaging of the near-field is
possible even without backward-wave materials (note that in these
cases the focusing of the propagating modes is not possible). This
phenomenon can be achieved with a bulk material slab having
negative permittivity or permeability~\cite{Fang}, or without any
bulk material by using planar sheets supporting surface
plasmons~\cite{Maslovski,Freire,Alitalo3}. Also devices which
operate in the ``canalization'' regime have been used successfully
to obtain subwavelength resolution~\cite{Belov,Ikonen}. Some of
the previous methods have also been suggested for use in the
optical region~\cite{Fang,Alu,Alitalo4}.

In this paper, we make a detailed study of a three-dimensional,
isotropic superlens based on loaded transmission-line networks.
This approach to superlens design that we use here was originally
proposed in~\cite{Alitalo1}. Here we confirm the isotropy of the
structure by studying the dependence of the dispersion on the
direction of propagation and verify the analytical design
equations presented in~\cite{Alitalo1} with full-wave simulations.
We also confirm that the impedance matching, which is essential
for operation of the device, is preserved for all directions of
propagation. The resolution enhancement capability of a
practically realizable, lossy lens is analytically studied using a
method presented in~\cite{Grbic2}. Although the lens inherently
achieves ideal operation at a single frequency only (the
dispersion curves of forward-wave and backward-wave materials
intersect at a single frequency point), we show that the
enhancement of the evanescent modes (which enables subwavelength
resolution) is possible in a small frequency band near the optimal
operation frequency.

\begin{figure}[h]
\centering \epsfig{file=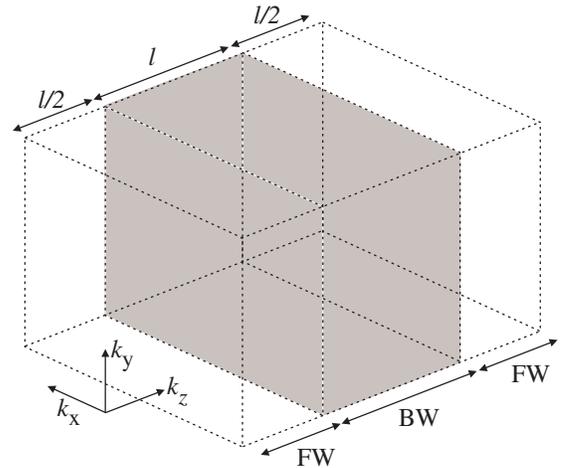, width=0.4\textwidth}
\caption{Structure of the superlens. If the source plane is at
$z=0$, the image plane is situated at $z=2l$.} \label{superlens}
\end{figure}

\section{The structure of the lens}

\begin{figure}[h]
\centering \epsfig{file=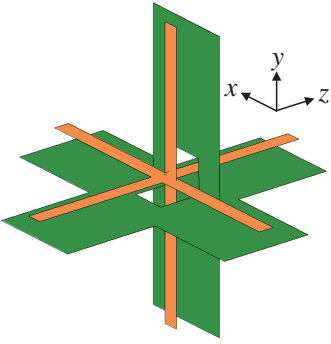, width=0.3\textwidth}
\caption{The unit cell of a three-dimensional, forward-wave
transmission-line network based on the microstrip technology (the
substrate is not shown for clarity). The backward-wave unit cell
is similar, but with lumped capacitors in each of the six branches
of the transmission lines (connected in series) and with a lumped
inductor, connected between the center node and the ground.}
\label{unit_cell}
\end{figure}

We study the superlens structure presented
in~\cite{Alitalo1,Alitalo2}. The superlens is a combination of two
types of transmission-line networks: a region with effectively
negative $\varepsilon$ and $\mu$ is sandwiched between two regions
possessing effectively positive $\varepsilon$ and $\mu$
(forward-wave regions), see Fig.~\ref{superlens}. As was
previously shown in~\cite{Alitalo1}, the transmission-line
networks are easy to realize using the microstrip technology and
we continue to use this approach in this paper (the design
equations can be applied to other types of transmission lines as
well, see~\cite{Alitalo1}). The forward-wave network has a unit
cell as shown in Fig.~\ref{unit_cell}. The unit cell of the
backward-wave network is otherwise similar to the one shown in
Fig.~\ref{unit_cell}, but it is loaded with lumped capacitors of
value $2C$ (in series with the microstrip lines) in all of the six
branches of the unit cell and an inductor of value $L$ is
connected from the center node of the microstrip line to the
ground. See~\cite{Alitalo2} for a representation of the unit cell
structure of both networks.

\section{Dispersion}

The dispersion equations for the forward-wave and backward-wave
networks have been derived in~\cite{Alitalo1} and they read (in
the lossless case):

\e \cos(q_{x})+\cos(q_{y})+\cos(q_{z})=\frac{1}{2j\omega LS_{\rm
BW}}-3\frac{K_{\rm BW}}{S_{\rm BW}}  \l{disp_BW} \f and \e
\cos(q_{x})+\cos(q_{y})+\cos(q_{z})=-3\frac{K_{\rm FW}}{S_{\rm
FW}},\l{disp_FW} \f where \e S_{\rm BW}=\frac{j\omega C}{(D_{\rm
t}+j\omega CB_{\rm t})(A_{\rm t}+j\omega CB_{\rm t})+\omega^2
C^2B_{\rm t}^2} \l{S_BW}, \f \e K_{\rm BW}=\frac{-(B_{\rm t}C_{\rm
t}-D_{\rm t}A_{\rm t})(A_{\rm t}+j\omega CB_{\rm t})}{[(D_{\rm
t}+j\omega CB_{\rm t})(A_{\rm t}+j\omega CB_{\rm
t})+\omega^{2}C^{2}B_{\rm t}^2]B_{\rm t}}-\frac{A_{\rm t}}{B_{\rm
t}}, \l{K_BW} \f \e S_{\rm FW}=\frac{1}{B_{\rm t}(A_{\rm t}+D_{\rm
t})} \l{S_FW}, \f \e K_{\rm FW}=-\frac{B_{\rm t}C_{\rm t}-D_{\rm
t}A_{\rm t}}{A_{\rm t}+D_{\rm t}}\frac{1}{B_{\rm t}}-\frac{A_{\rm
t}}{B_{\rm t}}, \l{K_FW} \f \e \left[
\begin{array}{ccc}
A_{\rm t} & B_{\rm t}  \\
C_{\rm t} & D_{\rm t}  \end{array} \right] = \left[
\begin{array}{ccc}
\cos(k_{\rm TL}d/2) & jZ_{0,\rm TL} \sin(k_{\rm TL}d/2)  \\
jZ_{0,\rm TL}^{-1} \sin(k_{\rm TL}d/2) & \cos(k_{\rm TL}d/2)
\end{array} \right]. \l{ABCD} \f

In \r{disp_BW}-\r{K_FW} the indices FW and BW correspond to the
dispersion equations of forward- and backward-wave networks,
respectively, and $q_{i}=k_{i}d$ (wavenumber normalized by the
period $d$), where $k_{i}$ is the wavenumber in the network along
axis $i$. In \r{ABCD} $k_{\rm TL}$ and $Z_{0,\rm TL}$ are the
wavenumber and impedance of the waves in the transmission lines,
respectively. Note that $Z_{0,\rm TL}$ is usually different for
the forward-wave and backward-wave networks in order to obtain
impedance matching of the networks~\cite{Alitalo1}.

\begin{figure}[h]
\centering \epsfig{file=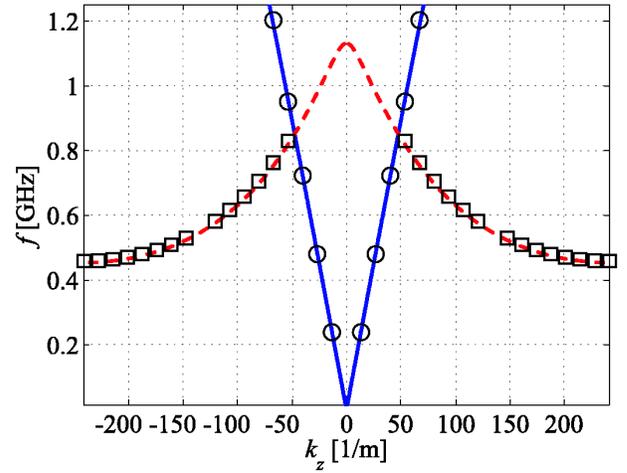, width=0.5\textwidth}
\caption{Dispersion curves for the forward-wave (solid line:
analytical; circles: HFSS) and backward-wave (dashed line:
analytical; squares: HFSS) networks. Propagation along the
$z$-axis is considered ($k_x = k_y = 0$). } \label{3D_BW-FW_z}
\end{figure}

The parameters of the structure that is studied here are the same
as in~\cite{Alitalo2}, see Table~\ref{table1} ($\varepsilon_{\rm
r}$ is the permittivity of the substrate of the microstrip lines).
Dispersion curves of the forward-wave and backward-wave networks
can be studied analytically using \r{disp_BW}-\r{ABCD}. The unit
cells of the both networks have also been simulated with Ansoft
HFSS full-wave simulator to obtain the dispersion curves for both
networks. See Fig.~\ref{3D_BW-FW_z} for the dispersion curves when
a planewave is considered ($k_x = k_y = 0$). The simulation
results agree very well for all axial directions (practically
identical plots). From Fig.~\ref{3D_BW-FW_z} we can conclude that
the optimal operation frequency of the superlens is $f=0.8513$~GHz
(the frequency at which the dispersion curves intersect) and at
that point the wavenumber has value $k_z\approx47.65$ 1/m.

\begin{table}[h]
\centering \caption{Parameters of the superlens structure.}
\label{table1}
\begin{tabular}{|c|c|c|c|c|c|} \hline

$d$ & $Z_{\rm 0,TL,FW}$ & $Z_{\rm 0,TL,BW}$ & $C$ & $L$ & $\varepsilon_{\rm r}$\\

\hline

13 mm & $66$ $\Omega$ & $89$ $\Omega$ & 3.3 pF & 6.8 nH & 2.33 \\

\hline
\end{tabular}
\end{table}

From \r{disp_BW} and \r{disp_FW} we clearly see that if we
consider diagonal propagation ($k_x$, $k_y$ and $k_z$ may all be
nonzero depending on the direction of propagation), this
introduces some anisotropy to the dispersion. To study this
effect, we have analyzed propagation in the structure in other
than the axial directions. It has been seen that for the
backward-wave network the isotropy is achieved in a large
bandwidth below and above the second stopband (for this example,
the isotropic region is approximately from 0.5~GHz to 2~GHz), and
for the forward-wave network at low frequencies (for this example,
below 2~GHz). The operation frequency of the designed lens is well
within this isotropic region for both networks. The optimal
operation frequency obtained here differs slightly from the one
previously presented for a similar structure~\cite{Alitalo2}. The
reason for this is that here we have assumed the effective
permittivity of the transmission lines to be equal to
$\varepsilon_{\rm r}$ (to simplify comparison between the
analytical and simulation results).

\begin{figure}[h]
\centering \epsfig{file=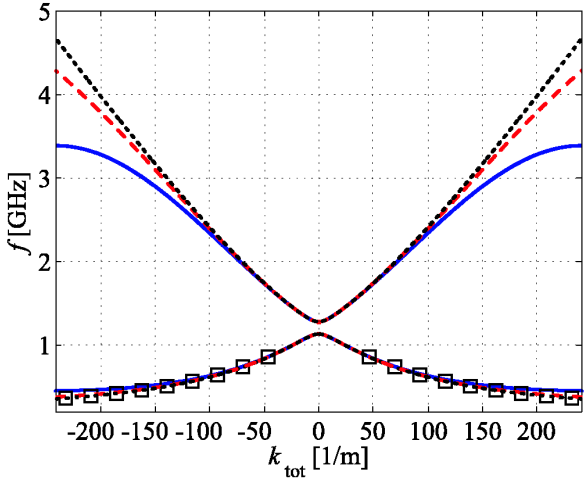, width=0.5\textwidth}
\caption{Dispersion curves for the backward-wave network. Solid
line: $k_x=k_y=0$; dashed line: $k_y=0$, $k_x=k_z$; dotted line:
$k_x=k_y=k_z$; squares: $k_x=k_y=k_z$ (HFSS).} \label{3D_BW_xyz}
\end{figure}

\begin{figure}[h]
\centering \epsfig{file=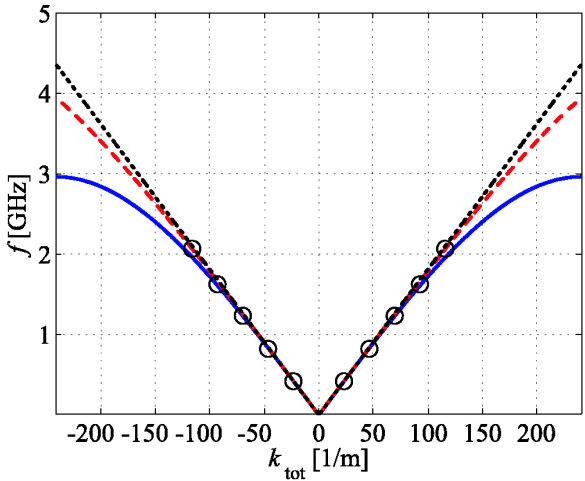, width=0.5\textwidth}
\caption{Dispersion curves for the forward-wave network. Solid
line: $k_x=k_y=0$; dashed line: $k_y=0$, $k_x=k_z$; dotted line:
$k_x=k_y=k_z$; circles: $k_x=k_y=k_z$ (HFSS).} \label{3D_FW_xyz}
\end{figure}

See Figs.~\ref{3D_BW_xyz} and \ref{3D_FW_xyz} for the results
considering different propagation directions. Note that for the
diagonal propagation, the dispersion curves extend to larger
values of $k_{\rm tot}=\sqrt{k_x^2+k_y^2+k_z^2}$ than it is shown
in Figs.~\ref{3D_BW_xyz} and \ref{3D_FW_xyz} (these regions are
not of interest for superlens operation). The HFSS simulation
results agree very well for all of the presented curves. For
clarity only the diagonal propagation corresponding to the case
with $k_x=k_y=k_z$ is presented (squares and circles).

\section{Impedance matching}

As was shown in~\cite{Alitalo1,Alitalo2}, the impedance matching
is crucial for the operation of the superlens. Here we use the
equations derived for the characteristic impedances of the
forward-wave and backward-wave networks~\cite{Alitalo1} to see if
the matching is preserved for all propagation directions. In the
following we study the matching analytically and tune the
impedances of the transmission lines slightly to obtain optimal
resolution performance (in~\cite{Alitalo2} and in the previous
section the impedance values were not ideal due to the fact that
the values were taken from an experimental prototype). It was
found that by changing the impedance of the forward-wave
transmission lines to $Z_{0,\rm TL,FW}=69.46$ $\Omega$, the
wavenumbers and the characteristic impedances of the networks can
be matched at the frequency $f=0.8513$~GHz. In the rest of this
paper, we use this impedance value and the other design
characteristics stay the same as shown in Table~\ref{table1}.

\begin{figure}[h]
\centering \epsfig{file=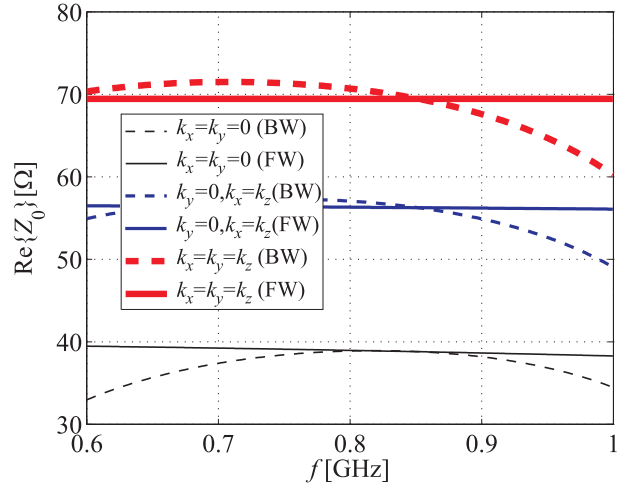, width=0.5\textwidth}
\caption{The characteristic impedances of forward-wave and
backward-wave networks for different propagation directions.}
\label{impedance}
\end{figure}

See Fig.~\ref{impedance} for the characteristic impedances of the
both networks for different propagation directions. Note that
because we are interested in the matching between the two
networks, the characteristic impedance is defined as the ratio of
the voltage and the $z$-component of the current (as was done
in~\cite{Alitalo1,Alitalo2}). We see that although the values of
the impedances change as the direction of propagation changes
(naturally, because the impedance depends on $k_z$), the matching
is preserved for different propagation directions at the operation
frequency ($f=0.8513$~GHz).

\section{Resolution characteristics}

\subsection{Resolution enhancement}

To evaluate the performance of the designed lens, we adopt the
same method of calculating the resolution enhancement as
in~\cite{Grbic2}, where the resolution enhancement was defined for
a two-dimensional (planar) lens as:

\e R_e=\frac{k_{t\rm,max}}{k_0}, \l{R_e} \f where $k_{t\rm,max}$
is the maximum transverse wavenumber that is transmitted from the
source plane to the image plane and $k_0$ is the maximum
transverse wavenumber corresponding to propagating modes
($k_0\approx47.65$ 1/m for the lens that we study here, as can be
seen from Fig.~\ref{3D_BW-FW_z}). Because we consider a
three-dimensional lens, the transverse wavenumber is now defined
as $k_t=\sqrt{k_x^2+k_y^2}$, see Fig.~\ref{superlens}.

In~\cite{Grbic2}, $k_{t\rm,max}$ was derived analytically from the
dimensions of the used superlens (taking into account the effect
of losses) as well as calculated from the optical transfer
function that was derived analytically and also measured. It was
concluded that a good approximation for $k_{t\rm,max}$ is the
value of $k_t$, at which the absolute value of the optical
transfer function drops to 0.5~\cite{Grbic2}. In the following, we
calculate the transmission coefficient of the lens studied in this
paper using the previously derived equations~\cite{Alitalo1}. From
the absolute value of the transmission coefficient (which
corresponds to the optical transfer function used
in~\cite{Grbic2}) we obtain $R_e$ by finding $k_{t\rm,max}$ from
the plotted curves as described above.

First, let us see how the thickness of the superlens affects the
resolution enhancement. We have calculated the resolution
enhancement for the superlens described in the previous sections,
taking into account realistic losses caused by the substrate and
by the lumped elements (loss tangent of the substrate is
$\tan\delta=0.0012$ and the quality factors of the capacitors and
inductors are 500 and 50, respectively)~\cite{Alitalo2}. In the
calculations, the losses can be taken into account by using
complex values for $C$ and $L$ and by replacing \r{ABCD} by

\e \left[
\begin{array}{ccc}
A_{\rm t} & B_{\rm t}  \\
C_{\rm t} & D_{\rm t}  \end{array} \right] = \left[
\begin{array}{ccc}
\cosh(\gamma d/2) & jZ_{0,\rm TL} \sinh(\gamma d/2)  \\
jZ_{0,\rm TL}^{-1} \sinh(\gamma d/2) & \cosh(\gamma d/2)
\end{array} \right], \l{ABCD_lossy} \f where \e \gamma=\frac{\pi\varepsilon_{\rm r}(\varepsilon_{\rm r}-1)\tan\delta}{ \sqrt{\varepsilon_{\rm r}}(\varepsilon_{\rm r}-1)\lambda_0 }+jk_{\rm TL}. \f

\begin{figure}[h]
\centering \epsfig{file=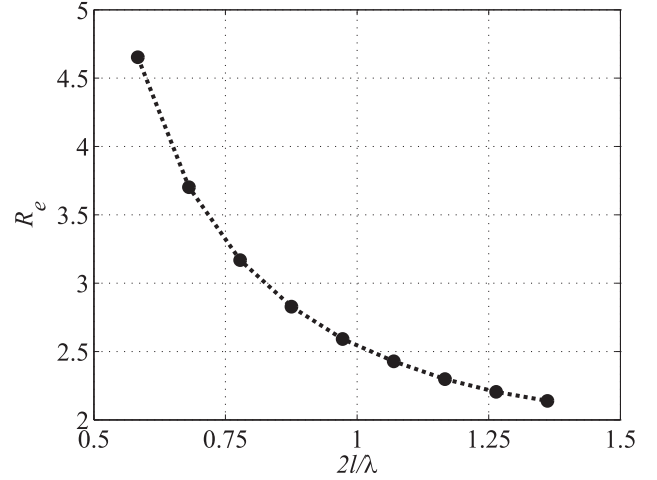, width=0.5\textwidth}
\caption{Resolution enhancement of the lens, as a function of the
thickness of the lens (distance between source and image) in
wavelengths at the optimal operation frequency ($f=0.8513$~GHz).
The larger dots show the points in which $R_e$ was analytically
calculated. Losses of the substrate and the lumped components are
taken into account.} \label{Re}
\end{figure}

See Fig.~\ref{Re} for the resolution enhancement as a function of
the thickness of the lens (here the thickness refers to the
distance from the source plane to the image plane, i.e., thickness
is equal to $2l$). Note that when we find $k_{t\rm,max}$ for
different thicknesses, the ``worst case'' is always used. Because
of the fact that the impedance values are different for different
directions of propagation (see Fig.~\ref{impedance}),
$k_{t\rm,max}$ is slightly different for various propagation
directions. As expected by the results of Fig.~\ref{impedance},
$|k_{t\rm,max}|$ is the smallest for the case when $k_x=k_y$. In
the calculation of Fig.~\ref{Re} and in section \textbf{V B} this
``worst case'' is used to calculate $R_e$.

\subsection{Bandwidth}

Ideal operation of the superlens described in this paper can be
achieved only at a single frequency, as can be seen from
Fig.~\ref{3D_BW-FW_z}. Nevertheless, it is clear that although a
small change in the frequency distorts the image seen in the image
plane, focusing of the propagating modes and enhancement of the
evanescent modes are still expected to happen in some small but
finite frequency band. Here we rely on the assumption that the
resolution enhancement can still be defined with \r{R_e}, i.e.,
the distortion of the image is not very dramatic and
$k_{t\rm,max}$ can still be defined as described above. In the
following, the operation band is defined as the frequency band
where $R_e>2$.

\begin{figure}[h]
\centering \epsfig{file=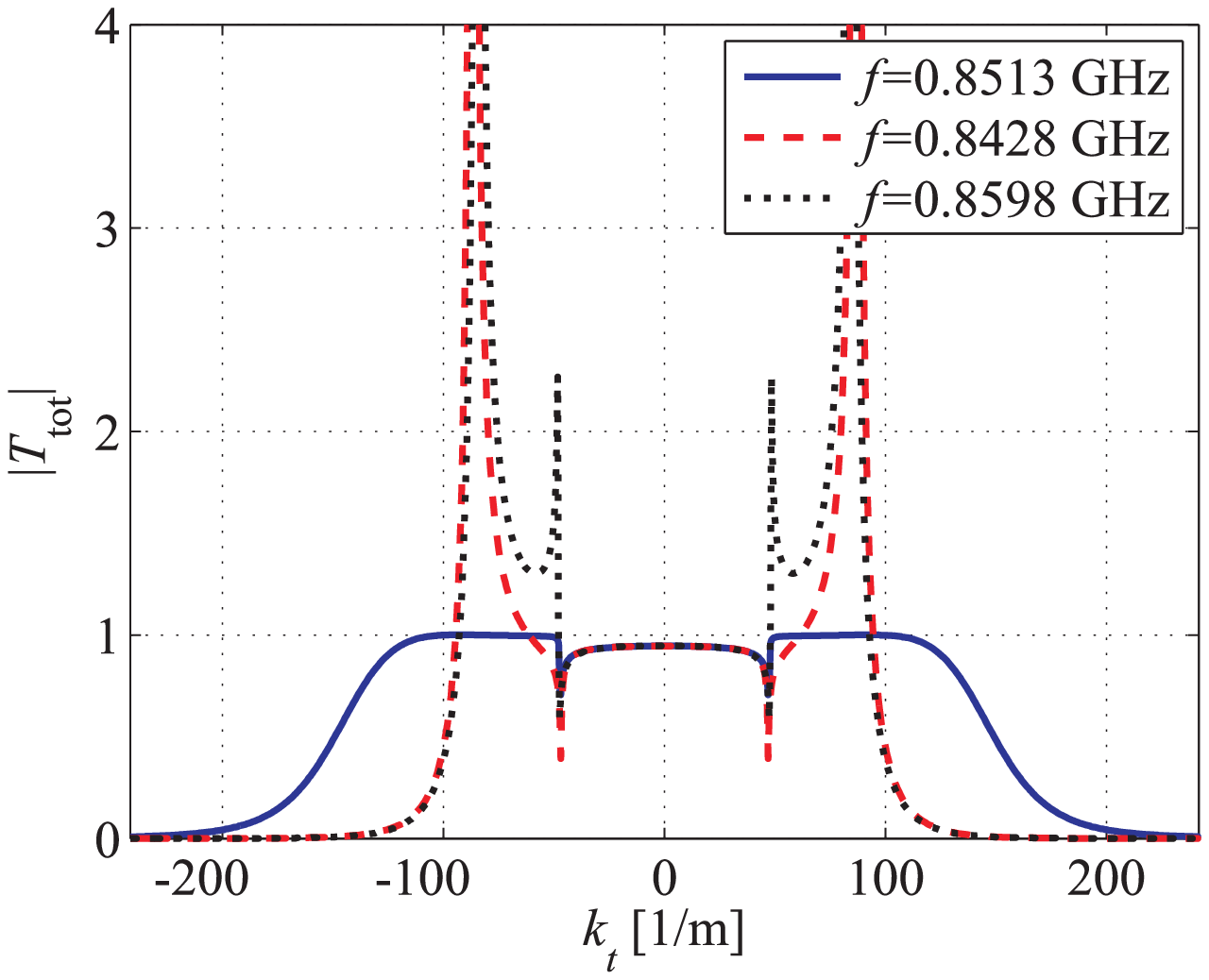, width=0.45\textwidth}
\caption{Transmission from the source plane to the image plane as
a function of the transverse wavenumber. Solid line:
$f=0.8513$~GHz; dashed line: $f=0.8482$~GHz; dotted line:
$f=0.8598$~GHz. The band where $R_e > 2$ is approximately 2~$\%$.
The distance between the source and image planes equals
$0.78\lambda$ at the center frequency.} \label{bandwidth}
\end{figure}

First, let us study a lens with the thickness of the backward-wave
network being 4 unit cells and the distance between the source and
image planes being 8 unit cells. In wavelengths this is
$0.78\lambda$ at the center frequency ($f=0.8513$~GHz), because
$\lambda=2\pi/k_{0}\approx 0.132$~m. See Fig.~\ref{bandwidth} for
a plot of the transmission coefficient (the absolute value). From
Fig.~\ref{bandwidth} we see that the bandwidth is approximately 2
percent.

\begin{figure}[h]
\centering \epsfig{file=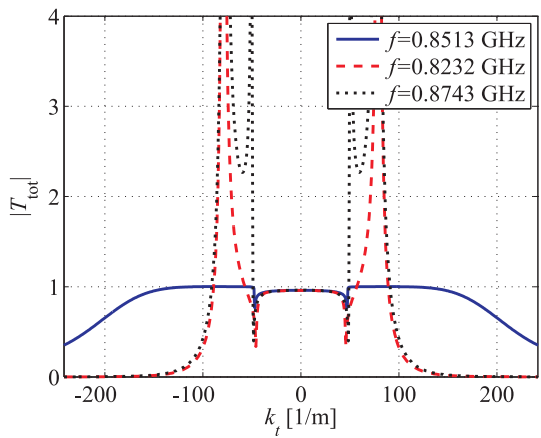,
width=0.45\textwidth} \caption{Transmission from the source plane
to the image plane as a function of the transverse wavenumber.
Solid line: $f=0.8513$~GHz; dashed line: $f=0.8232$~GHz; dotted
line: $f=0.8743$~GHz. The band where $R_e > 2$ is approximately
6~$\%$. The distance between the source and image planes equals
$0.58\lambda$ at the center frequency.} \label{bandwidth2}
\end{figure}

Next, we make the lens thinner to see how this affects the
bandwidth. Now the thickness of the backward-wave network is 3
unit cells and the distance between the source and image planes is
6 unit cells. In wavelengths this is $0.58\lambda$ at the center
frequency. See Fig.~\ref{bandwidth2} for the absolute value of the
transmission coefficient corresponding to this case. From
Fig.~\ref{bandwidth2} we can conclude that the bandwidth is
approximately 6 percent.

\section{Conclusions}

We have shown that a three-dimensional isotropic transmission-line
network can be designed in such a way that the effective
permittivity and permeability of the network are negative (a
backward-wave material). When combined with a transmission-line
network with positive effective permittivity and permeability, the
resulting device (superlens) can achieve subwavelength resolution
in a small frequency band. In this paper we have verified the
previously derived dispersion equations by full-wave simulations
and have shown that the designed structure is isotropic in all
propagation directions (not just along the three orthogonal ones).
We have also confirmed that impedance matching of the two types of
networks is possible for an arbitrary direction of propagation. We
have analytically studied the effect of losses and the physical
size on the resolution and bandwidth characteristics of the
designed lens. When high-quality, low-loss components and
materials are used, the designed lens can achieve substantial
resolution enhancement in a relative bandwidth of a few percents,
with the thickness of the lens being of the order of one
wavelength.

\section*{Acknowledgements}

This work has been partially funded by the Academy of Finland and
TEKES through the Center-of-Excellence program. The authors wish
to thank Mr. T. Kiuru and Mr. O. Luukkonen for helpful discussions
regarding the simulation software. Pekka Alitalo wishes to thank
the Graduate School in Electronics, Telecommunications and
Automation (GETA) and the Nokia Foundation for financial support.

\end{document}